\documentclass[twocolumn,times]{aastex7}
\usepackage{graphicx}
\usepackage{subcaption}
\usepackage{fancyhdr}

\begin{document}

\mbox{FERMILAB-PUB-25-0541-CSAID-PPD}

\title{A Joint Search for the Electromagnetic Counterpart to the Gravitational-Wave Binary Black-Hole Merger Candidate S250328ae with the Dark Energy Camera and the Prime Focus Spectrograph\footnote{Based on data collected at Subaru Telescope, which is operated by the National Astronomical Observatory of Japan.}}

\begin{NoHyper}
\correspondingauthor{haibin.zhang@nao.ac.jp}
\end{NoHyper}

\author[orcid=0000-0003-2273-9415,gname=Haibin, sname=Zhang]{Haibin Zhang} 
\affiliation{National Astronomical Observatory of Japan, 2-21-1 Osawa, Mitaka, Tokyo 181-8588, Japan}
\email{haibin.zhang@nao.ac.jp}

\author[orcid=0000-0001-6402-1415,gname=Mitsuru, sname=Kokubo]{Mitsuru Kokubo}
\affiliation{National Astronomical Observatory of Japan, 2-21-1 Osawa, Mitaka, Tokyo 181-8588, Japan}
\email{mitsuru.kokubo@nao.ac.jp}

\author[orcid=0000-0002-9514-7245,gname=Sean,sname=MacBride]{Sean MacBride}
\affiliation{Physik-Institut, University of Zurich, Winterthurerstrasse  190, 8057 Zurich, Switzerland}
\email{sean.macbride@physik.uzh.ch} 

\author[orcid=0000-0002-4529-1505,gname=Isaac,sname=McMahon]{Isaac McMahon}
\affiliation{Physik-Institut, University of Zurich, Winterthurerstrasse  190, 8057 Zurich, Switzerland}
\email{isaac.mcmahon@physik.uzh.ch} 

\author[orcid=0000-0001-8537-3153,gname=Nozomu, sname=Tominaga]{Nozomu Tominaga}
\affiliation{National Astronomical Observatory of Japan, 2-21-1 Osawa, Mitaka, Tokyo 181-8588, Japan}
\affiliation{Astronomical Science Program, The Graduate University for Advanced Studies (SOKENDAI), 2-21-1 Osawa, Mitaka, Tokyo 181-8588, Japan}
\affiliation{Department of Physics, Konan University, 8-9-1 Okamoto, Kobe, Hyogo 658-8501, Japan}
\email{nozomu.tominaga@nao.ac.jp}

\author[orcid=0000-0001-6161-8988,gname=Yousuke, sname=Utsumi]{Yousuke Utsumi} 
\affiliation{National Astronomical Observatory of Japan, 2-21-1 Osawa, Mitaka, Tokyo 181-8588, Japan}
\affiliation{Vera C.\ Rubin Observatory, Avenida Juan Cisternas \#1500, La Serena, Chile}
\email{yousuke.utsumi@nao.ac.jp}

\author[orcid=0000-0002-9948-1646,gname=Michitoshi,sname=Yoshida]{Michitoshi Yoshida}
\affiliation{National Astronomical Observatory of Japan, 2-21-1 Osawa, Mitaka, Tokyo 181-8588, Japan}
\email{michitoshi.yoshida@nao.ac.jp}

\author[orcid=0000-0001-7449-4814,gname=Tomoki,sname=Morokuma]{Tomoki Morokuma}
\affiliation{Astronomy Research Center, Chiba Institute of Technology, 2-17-1 Tsudanuma, Narashino, Chiba 275-0016, Japan}
\email{tmorokuma@perc.it-chiba.ac.jp}

\author[orcid=0000-0001-8253-6850,gname=Masaomi,sname=Tanaka]{Masaomi Tanaka}
\affiliation{Astronomical Institute, Tohoku University, Aoba, Sendai 980-8578, Japan}
\email{masaomi.tanaka@astr.tohoku.ac.jp}

\author[orcid=0000-0002-5756-067X,gname=Akira,sname=Arai]{Akira Arai}
\affiliation{National Astronomical Observatory of Japan, 2-21-1 Osawa, Mitaka, Tokyo 181-8588, Japan}
\email{arai@naoj.org}

\author[orcid=0000-0001-7759-6410,gname=Wanqiu,sname=He]{Wanqiu He}
\affiliation{National Astronomical Observatory of Japan, 2-21-1 Osawa, Mitaka, Tokyo 181-8588, Japan}
\email{wanqiu.he@nao.ac.jp}

\author[orcid=0000-0003-2541-1052,gname=Yuki,sname=Moritani]{Yuki Moritani}
\affiliation{National Astronomical Observatory of Japan, 2-21-1 Osawa, Mitaka, Tokyo 181-8588, Japan}
\email{moritani@naoj.org}

\author[orcid=0000-0003-3228-7264,gname=Masato,sname=Onodera]{Masato Onodera}
\affiliation{National Astronomical Observatory of Japan, 2-21-1 Osawa, Mitaka, Tokyo 181-8588, Japan}
\email{monodera@naoj.org}

\author[orcid=0000-0002-8569-7243,gname=Vera Maria,sname=Passegger]{Vera Maria Passegger}
\affiliation{National Astronomical Observatory of Japan, 2-21-1 Osawa, Mitaka, Tokyo 181-8588, Japan}
\affiliation{Hamburger Sternwarte, Gojenbergsweg 112, 21029 Hamburg, Germany}
\email{vmpas@naoj.org}

\author[orcid=0000-0002-4937-4738,gname=Ichi,sname=Tanaka]{Ichi Tanaka}
\affiliation{National Astronomical Observatory of Japan, 2-21-1 Osawa, Mitaka, Tokyo 181-8588, Japan}
\email{ichi@naoj.org}

\author[orcid=0000-0001-6229-4858,gname=Kiyoto,sname=Yabe]{Kiyoto Yabe}
\affiliation{National Astronomical Observatory of Japan, 2-21-1 Osawa, Mitaka, Tokyo 181-8588, Japan}
\email{kiyoyabe@naoj.org}

\author[orcid=0009-0006-5765-1607,gname=Lillian,sname=Joseph]{Lillian Joseph}
\affiliation{Benedictine University, Department of Physics, 5700 College Road, Lisle, IL 60532, USA}
\email{liljo@sas.upenn.edu} 

\author[orcid=0009-0004-8314-0427,gname=Simran,sname=Kaur]{Simran Kaur}
\affiliation{Department of Physics, University of Michigan, Ann Arbor, MI 48109, USA}
\email{simrankj@umich.edu} 

\author[orcid=0009-0007-4271-6444,gname=Hemanth,sname=Bommireddy]{Hemanth Bommireddy}
\affiliation{ Department of Astronomy, Universidad de Chile, Camino el Observatorio 1515, Las Condes, Santiago, Chile}
\email{hemanth@das.uchile.cl}

\author[orcid=0000-0001-5399-0114,gname=Nora,sname=Sherman]{Nora Sherman}
\affiliation{Institute for Astrophysical Research, Boston University, 725 Commonwealth Avenue, Boston, MA 02215, USA}
\email{norafsherman@att.net} 

\author[orcid=0000-0001-6718-2978,gname=Kenneth,sname=Herner]{Kenneth Herner}
\affiliation{Fermi National Accelerator Laboratory, Kirk Road and Pine Street, Batavia, IL 60510, USA}
\email{kherner@fnal.gov} 

\author[orcid=0000-0002-8357-7467,gname=H. Thomas,sname=Diehl]{H. Thomas Diehl}
\affiliation{Fermi National Accelerator Laboratory, Kirk Road and Pine Street, Batavia, IL 60510, USA}
\email{diehl@fnal.gov} 

\author[orcid=0000-0001-6082-8529,gname=Marcelle,sname=Soares-Santos]{Marcelle Soares-Santos}
\affiliation{Physik-Institut, University of Zurich, Winterthurerstrasse  190, 8057 Zurich, Switzerland}
\email{marcelle@physik.uzh.ch} 

\collaboration{all}{Japanese Collaboration for Gravitational-Wave Electro-Magnetic Follow-up (J-GEM)}

\collaboration{all}{The Dark Energy Survey Gravitational Wave Collaboration}

\begin{abstract}

The first detection of an optical counterpart to a gravitational wave signal revealed that collaborative efforts between instruments with different specializations provide a unique opportunity to acquire impactful multi-messenger data. We present results of such a joint search with the Dark Energy Camera (DECam) and Prime Focus Spectrograph (PFS) for the optical counterpart of the LIGO-Virgo-KAGRA event S250328ae, a binary black hole merger candidate of high significance detected at a distance of 511$\pm$82 Mpc and localized within an area of 3 (15) square degrees at 50\% (90\%) confidence. We observed the 90\% confidence area with DECam and identified 36 high-confidence transient candidates after image processing, candidate selection, and candidate vetting. We observed with PFS to obtain optical spectra of DECam candidates, Swift-XRT candidates, and potential host galaxies of S250328ae. In total, 3897 targets were observed by seven pointings covering $\sim 50\%$ of the 90\% confidence area. After template fitting and visual inspection, we identified 12 SNe, 159 QSOs, 2975 galaxies, and 131 stars. With the joint observations of DECam and PFS, we found variability in 12 SNe, 139 QSOs, 37 galaxies, and 2 stars. We do not identify any confident optical counterparts, though the association is not ruled out for three variable candidates that are not observed by PFS and 6 QSO candidates without clear variability if the optical counterpart of S250328ae is faint. Despite the lack of confident optical counterparts, this paper serves as a framework for future collaborations between wide-field imagers and multi-object spectrographs to maximize multi-messenger analyses.
\end{abstract}



\section{Introduction}\label{sec:intro}

Since the detection of the Binary Neutron Star (BNS) gravitational wave event GW170817 \citep{GW170817} and its associated optical counterpart kilonova through a global search effort \citep{Abbott2017_mm17, SoaresSantos2017, Utsumi2017, Coulter2017, Andreoni2017, Evans2017, Valenti2017, Shappee2017,Tominaga2018}, significant attempts have been made across many institutions and instruments to discover more electromagnetic counterparts to gravitational wave events. As demonstrated with GW170817, the detection of a counterpart can open the door to making cosmological measurements \citep{Schutz1986, Abbott2017_siren}, studying kilonova ejecta dynamics \citep{Finstad2018, Makhathini2021}, constraining the equation of state of neutron stars \citep{Abbott2018_nsEOS}, among other studies. This growing field of multi-messenger astrophysics has not yet yielded any additional confirmed optical counterpart detections after GW170817.

Despite optimistic early estimates of BNS merger rates, only two such events have ever been detected, while many more Binary Black Hole (BBH) events have been detected to date \citep{Abbott2023_rates_pop}. Electromagnetic counterparts from BNS events are the most informative and well-studied, but BBH and Neutron Star-Black Hole (NSBH) mergers are also expected to produce a detectable counterpart in favorable conditions \citep{corsi2024multimessengerastrophysicsblackholes}. Leading models predict that electromagnetic emission from a BBH event may be produced in the dense environments of the accretion disks of Active Galactic Nuclei (AGN) through various astrophysical processes \citep{AGN_BBH, Kimura2021, RodrguezRamrez2025, Gr_bner_2020, Bartos2017, McKernan2019, Tagawa2024}. This emission is inherently more difficult to detect due to the intrinsic variability of the host AGN. Moreover, many events will not produce any detectable emission due to the strong coupling between the initial conditions and the resulting luminosity. No confirmed electromagnetic counterpart to a BBH event has been observed yet, despite numerous attempts \citep{Graham_2020, Kim_2021, darc2025longtermopticalfollows231206cc}.

The LIGO-Virgo-KAGRA Collaboration (LVK) is in the middle of its fourth observing run (O4), which began in May 2023 and is currently expected to end in November 2025. The collaboration has detected nearly 300 high-significance gravitational wave events in total, of which the majority have been BBH events. Upon detection of a new gravitational wave event, LVK immediately sends a notice to the wider scientific community via the General Coordinates Network (GCN) time-domain alert system.

The Dark Energy Camera (DECam) is an instrument on the Blanco 4-Meter Telescope at the Cerro Tololo Inter-American Observatory (CTIO) in Chile \citep{Flaugher_2015}, built to support the Dark Energy Survey (DES) \citep{2005astro.ph.10346T, Abbott_2018}. DECam is an ideal EM counterpart discovery and characterization machine due to a $3\deg^2$ FOV and a single-image $10\sigma$ limiting magnitude $\sim23$ in \textit{r}-band \citep{Abbott_2021, Morganson2018}. DES conducts gravitational wave electromagnetic counterpart search campaigns through the DES Gravitational Wave (DESGW) program \citep{Herner2020}. DESGW was among the first groups to confirm the detection of the optical counterpart kilonova to GW170817 \citep{SoaresSantos2017}. DESGW has also previously conducted search campaigns on a number of BBH events \citep{SoaresSantos2016_GW150914, Cowperthwaite2016, Doctor2019, Morgan2020, Garcia2020}. Other collaborations have also used DECam to conduct search campaigns, such as GROWTH \citep{Goldstein2019, Andreoni2019_S190510g, Andreoni2020_S190814bv} and GW-MMADS \citep{Cabrera2024}.

The Japanese collaboration for Gravitational wave ElectroMagnetic follow-up (J-GEM) is a group searching for electromagnetic counterparts of GW events utilizing telescopes in Japan and other countries (e.g., US, China, New Zealand, Chile, and South Africa). The J-GEM team has carried out and reported follow-up observations of a number of GW events from the first event GW150914 (\citealt{2017PASJ...69....9Y,Morokuma2016,Utsumi2017,Tominaga2018,Sasada2021,Ohgami2021,Ohgami2023}). The uniqueness of AT2017gfo as an optical counterpart of GW170817 is only confirmed by the wide-field imaging follow-up observations by the J-GEM with Subaru/Hyper Suprime-Cam (HSC) \citep{Tominaga2018} and the DESGW with DECam \citep{SoaresSantos2017}.

The Subaru/Prime Focus Spectrograph (PFS) is a fiber spectrograph installed at the 8.2-meter Subaru telescope atop Maunakea in Hawaii, USA \citep{Sugai2015,Tamura2024}. PFS can observe $\sim 2400$ targets simultaneously within a $\sim 1.25 \deg^2$ field of view (FoV). The fiber diameter is $\sim 1.1$ arcsec at the FoV center and $\sim 1.0$ arcsec at the edge. The PFS contains four spectrographs, each of which consists of three spectral arms (blue, red, and NIR) with a wavelength coverage of 380-650, 630-970, and 940-1260 nm, respectively. The resolving power is $\sim 2500-4500$ depending on the wavelength. Alternatively, there is also a medium resolution mode for the red arm with a resolving power of $\sim 5500$ and a wavelength coverage of 710-885 nm. The large field of view and numerous fibers make PFS a powerful instrument for electromagnetic counterpart search of GW events.


Here, we present the results of a joint search for the gravitational wave BBH event S250328ae with DECam and PFS. In section \ref{sec:Data}, we review the data reported from four experiments: LVK, DESGW, PFS, and Swift's X-Ray Telescope (Swift-XRT). In section \ref{sec:Methods}, we review the methods for processing and analyzing data from the DESGW and PFS experiments. We present our results and discussion in section \ref{sec:results}. Finally, we summarize this study in section \ref{sec:summary}. Throughout this paper, we use the AB magnitudes \citep{Oke1983} and assume a $\Lambda$CDM cosmology with $\Omega_b=0.05$, $\Omega_c=0.24$, $\Omega_\Lambda=0.71$, and $h=0.69$ \citep{Hinshaw2013}. 

\section{Data} \label{sec:Data}

\subsection{Gravitational Wave Detection}\label{subsec:GWData}

The LVK collaboration reported the compact binary merger candidate S250328ae on 2025-03-28 05:40:27.419 UTC during real-time analysis of data from the LIGO Hanford (H1), LIGO Livingston (L1), and Virgo (V1) detectors \citep{LVKGCN_1,GraceDB_2025}. The candidate was independently detected by multiple LVK real-time analysis pipelines, including \texttt{gstlal}, \texttt{pycbc}, \texttt{cWB}, \texttt{cWB BBH}, \texttt{MBTA}, \texttt{Mly}, and \texttt{spiir} \citep{GstLALPipeline_0, GstLALPipeline_1, PyCBCLivePipe, cwbPipeline,cwbbbhPipeline,MBTAPipeline, MLyPipe, SPIIRPipe, LastGWPipe}. The event was reported with a false alarm rate (FAR) of $3.2\times 10^{-10}$ Hz or 1 per 100 years from \texttt{pycbc}.

The \texttt{RapidPE-RIFT} pipeline \citep{rapidPEPipe} initially classified the merger as a BBH merger with $>99\%$ probability. The likelihood that either compact object lay in the mass gap ($3-5M_\odot$) was estimated at $7\%$, while the probabilities of containing a neutron star (HasNS) or forming a remnant disk (HasRemnant) were both $<1\%$, based on a range of neutron star equations of state \citep{Chatterjee_2020}.

While sky localization was initially performed using \texttt{BAYESTAR} \citep{Bayestar}, a subsequent skymap produced by \texttt{Bilby} parameter estimation \citep{bilby} refined the 90\% confidence area to 15 $\deg^2$ \citep{LVKGCN_2}. The luminosity distance from \texttt{Bilby}, marginalized over the entire sky, was reported as $511\pm82$ Mpc ($z \sim 0.11 \pm 0.02$). We used this skymap to plan our observations and analysis.

\subsection{Swift-XRT Observations} \label{subsec:SwiftData}

The Neil Gehrels Swift Observatory conducted follow-up observations of S250328ae, targeting regions of interest informed by the \texttt{BAYESTAR} 3D localization map convolved with the 2MASS Photometric Redshift (2MPZ) galaxy catalog \citep{Bilicki_2014}. A total of 97 pointings were executed, spanning from $\sim1$ hour to $\sim10.5$ hours after the LVK trigger and covering 5.3 $\deg^2$ on the sky. This corresponds to 68\% of the \texttt{BAYESTAR} confidence area, or 72\% after galaxy convolution \citep{SwiftFollowup}.

In total, 34 previously uncatalogued X-ray sources were detected. All of these sources were classified as rank 3 detections, meaning they were new sources but did not exceed archival flux limits and were thus unlikely to be associated with S250328ae. No sources of higher interest were discovered. These findings were reported in GCN 39972 \citep{GCN39972}.

\subsection{DESGW Observations}\label{subsec:DECamData}

\begin{figure}[b]
    \centering
    \includegraphics[width=0.9\linewidth]{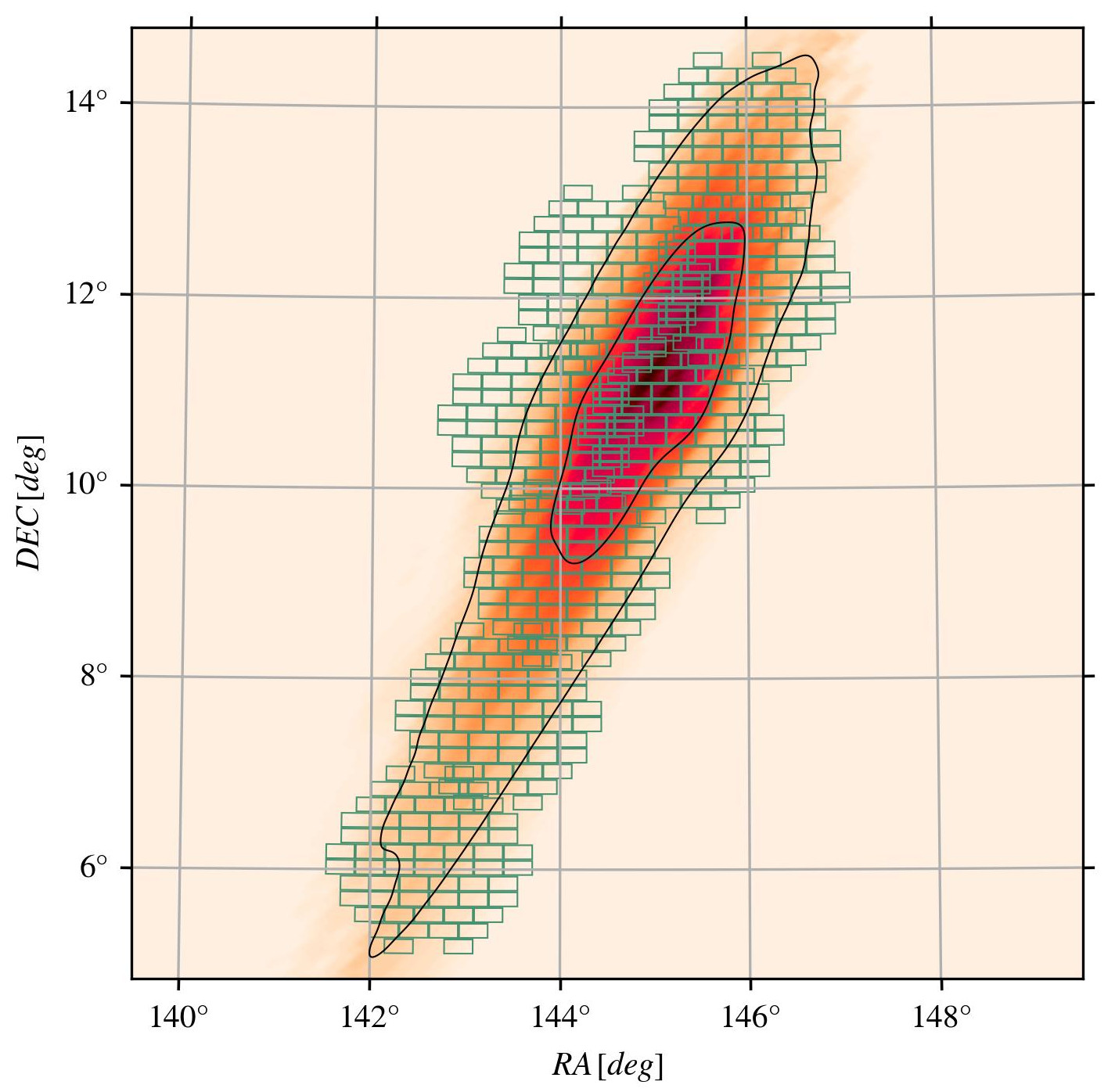}
    \caption{The \texttt{Bilby} skymap for S250328ae colored by the probability of originating in each pixel. 50\% and 90\% contours are shown in black. The eight pointings used by DESGW are shown in green, with the focal plane of DECam overlaid.}
    \label{fig:DESGW_obsPlan}
\end{figure}

\begin{figure}[b]
    \centering
    \includegraphics[width=\linewidth]{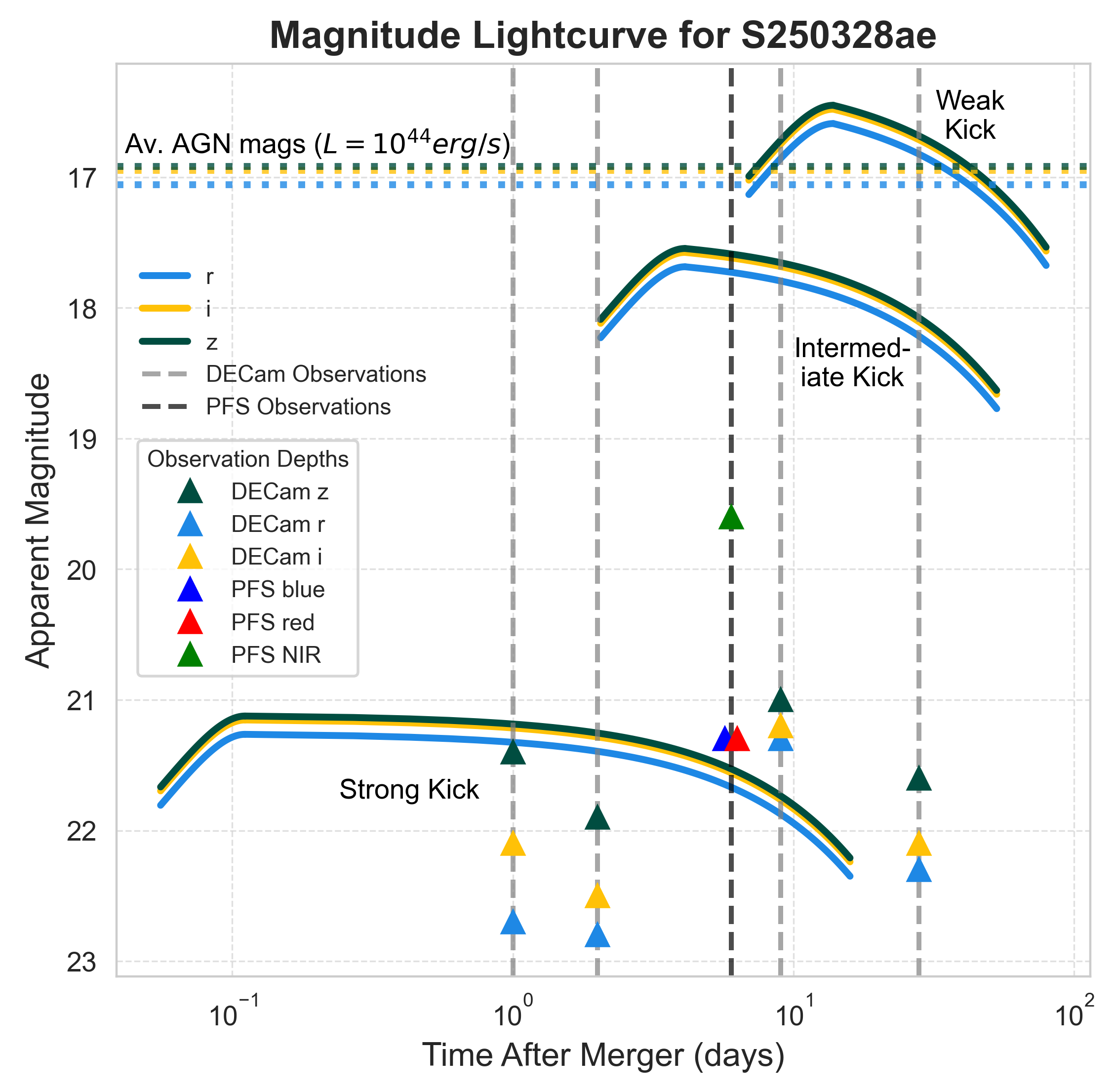}
    \caption{Simulated light curves for S250328ae in DECam \textit{riz} bands, following eq. 5 of \cite{Graham_2020}, compared to a typical AGN Luminosity. The different light curves correspond to flares produced by strong (1000 km/s), intermediate (300 km/s), and weak (200 km/s) kick velocities imparted to the BBH merger remnant within an AGN disc. The $10\sigma$ ($5\sigma$) limiting magnitudes of DECam (PFS) from table \ref{tab:ObservationTable} indicate that PFS could detect a BBH with intermediate and weak kick velocities, while DECam could observe all three BBH models, except the strong kick model on the third and fourth epochs of observation.}
    \label{fig:DESGW_BBHLC}
\end{figure}

DESGW was notified of S250328ae on 2025-03-28 05:47:08 UTC through the DESGW real-time monitoring code, Main-Injector \citep{MIPaper}, and officially triggered Target-of-Opportunity (ToO) at 2025-03-28 15:38:00 UTC. The observing plan was created by Main-Injector, which uses a set of predetermined sky tilings and ranks them by skymap probability coverage. The observing plan for S250328ae consists of 8 pointings in each of the DECam \textit{r}, \textit{i}, and \textit{z} bands, covering more than 90\% of the 90\% confidence area (see Figure \ref{fig:DESGW_obsPlan}). An exposure time of 90s was chosen to probe the necessary depth of the probability volume, resulting in a total exposure time of $\sim36$ minutes per night.

We supported our decision to trigger a follow-up of this event by estimating the chirp mass of the binary using machine learning techniques. We employ an Ordinal Class Probability Density Function algorithm \citep{Rau2015_OCP} adapted for gravitational wave data. Using \texttt{BAYESTAR}, we simulate 368,000 events with the published LVK detector noise curves for O4. The simulated skymaps and chirp masses serve as training data for a random forest classifier, provided by the \texttt{sklearn} python package \citep{sklearn}. Using this classifier to predict chirp masses from skymaps, we use the initial \texttt{BAYESTAR} skymap to estimate a chirp mass of $12^{+9}_{-6}\;M_\odot$ for S250328ae.

We observed four epochs listed in Table \ref{tab:ObservationTable} with identical observing plans. Epoch timing was scheduled to maximize the probability of characterizing the EM counterpart of a BBH merger for a range of remnant kick velocities following \cite{Graham_2020} and \cite{AGN_BBH} (see Figure \ref{fig:DESGW_BBHLC}). Additionally, DECam was scheduled for engineering and maintenance for ten nights in April, which added additional constraints to epoch timing.

\subsection{PFS Observations}\label{subsec:PFS_observations}

\begin{figure}[b]
    \centering
    \includegraphics[width=0.95\linewidth]{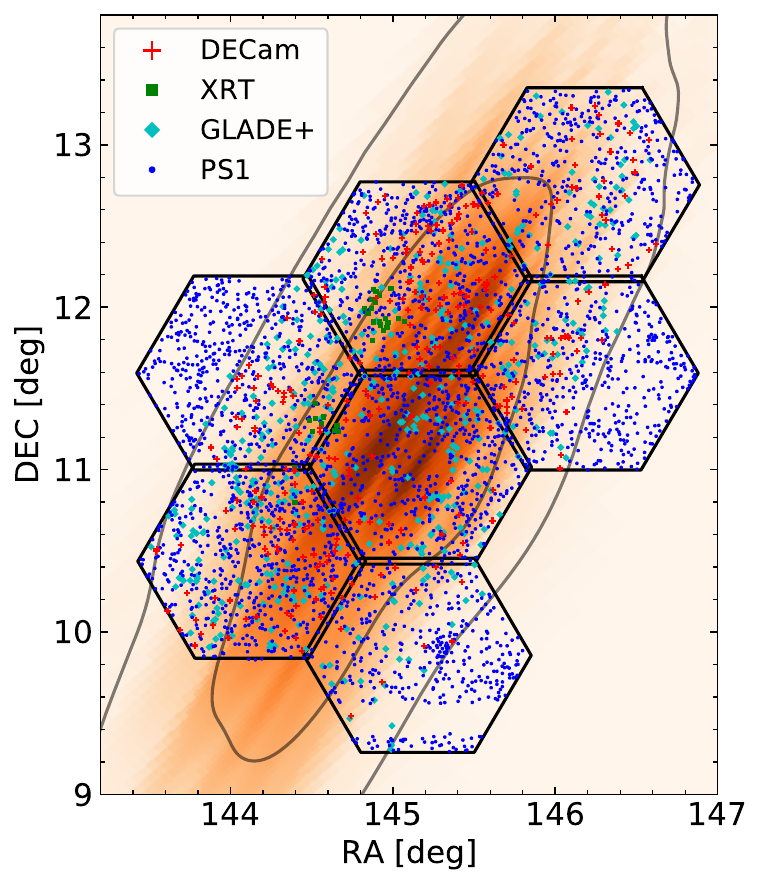}
    \caption{Same as Figure \ref{fig:DESGW_obsPlan}, but for Subaru/PFS observations. The grey contours correspond to the 50\% and 90\% probability regions of S250328ae. The red, green, cyan, and blue markers are DECam, Swift-XRT, GLADE+, and PS1 targets we observed, respectively. The black hexagons show the seven PFS pointings. The exposure time is 900 seconds for the bottom pointing and 1800 seconds for the other six pointings. Another PFS pointing at the bottom left corner (not shown in this figure) was planned but lost due to weather conditions.}
    \label{fig:PFS_pointings}
\end{figure}

\begin{table*}[t]
\centering
\hskip-2cm\begin{tabular}{|c|c|c|}
    \hline
    Observation Night & Facility & Band: Limiting Magnitude $m_{lim}$ \\
    \hline
    2025-03-29 & DECam (Blanco) & $z: 21.4,\;\; i: 22.1,\;\; r:22.7$ \\
    \hline
    2025-03-30 & DECam (Blanco) & $z: 21.9,\;\; i: 22.5,\;\; r:22.8$ \\
    \hline
    2025-04-03 & PFS (Subaru) & $blue: 21.3,\;\; red: 21.3,\;\; NIR:19.6$\\
    \hline
    2025-04-06 & DECam (Blanco) & $z: 21.0,\;\; i: 21.2,\;\; r:21.3$ \\
    \hline
    2025-04-25 & DECam (Blanco) & $z: 21.6,\;\; i: 22.1,\;\; r:22.3$ \\
    \hline
\end{tabular}
\caption{Observation dates and notes for DECam and PFS observations of S250328ae. DECam $10\sigma$ limiting magnitudes for the exposures are estimated using \cite{T_eff}. PFS limiting magnitudes are estimated for continuum using $5\sigma$ median noise of single pixels (pixel size $\sim 0.8$ {\AA}) in blue, red, and NIR spectral arms, respectively.}
\label{tab:ObservationTable}
\end{table*}

We triggered a ToO spectroscopic observation with PFS to search for the EM counterpart to S250328ae on 2025-04-02 UT. The observation was carried out from 05:11 to 10:20 UT on 2025-04-03. The PFS was set to the low resolution mode for full wavelength coverage. We observed seven pointings covering  $\gtrsim 90\%$ of the 50\% (and $\sim 50\%$ of the 90\%) confidence area of S250328ae, as shown in Figure \ref{fig:PFS_pointings}. The weather conditions were mostly good with a typical seeing size of $\lesssim 0.8$ arcsec, although the transparency occasionally dropped to $\lesssim 0.6$ due to passing clouds. The total on-source exposure time was $4 \times 450 = 1800$ seconds for six pointings and $2 \times 450 = 900$ seconds for one pointing that was shortened due to weather conditions. 

Based on the DESGW observation and archival data, we selected science targets for PFS follow-up observations. In the order from high priority to low priority, our science targets included DESGW transient candidates identified during the first two epochs, X-ray sources reported in GCN 39972 \citep{GCN39972}, potential host galaxies from the GLADE+ catalog \citep{Dalya2022}, and potential host galaxies from the PS1-STRM catalog \citep{Beck2021} with photometric redshift (photo-z) errors of $<0.2$. The spec-z or photo-z of selected potential host galaxies overlapped with the 90\% confidence interval of S250328ae redshift. The DESGW transient candidates were further divided into those reported in GCN 39934 and 39992 (Priority 0; highest priority), detected on both first and second DECam epochs (Priority 1), and detected on either DECam epochs (Priority 2). The priorities of our targets were used for science fiber allocation, and targets with a higher priority were more likely to be observed. In total, 3897 science targets were selected for our PFS observations based on the PFS planner. 

\section{Methods} \label{sec:Methods}

\subsection{DESGW Image Processing and Candidate Vetting} \label{subsec:DESGW candidate vetting}

We process images using the DESGW difference imaging pipeline. The DESGW difference imaging pipeline uses public DECam images as templates \citep{Morganson2018,2020EPJWC.24501008H}. We then run \texttt{SExtractor} \citep{SourceExtractor} on each image to produce a list of sources in the image, which enables \texttt{scamp} \citep{scamp} to perform astrometric calibration. Template images are subtracted from search images using the \texttt{diffimg} code developed for DES supernova searches \citep{Kessler_2015}. This results in a list of transient candidates that are found in the subtracted images. 

We use the \texttt{autoscan} machine learning code \citep{AutoscanPaper} to reject subtraction artifacts. We match our candidates against the ALLWISE, Milliquas, Quaia, and LQAC-6 AGN catalogs \citep{ALLWISE, Milliquas, Quaia, LQAC-6} within the LVK localization volume to search for possible known AGN hosts. Additionally, we attempt to match our candidates to any candidates that were reported by concurrent follow-up efforts of S250328ae.

We employ a number of criteria to filter only for candidates of immediate interest. First, we require that each candidate must have at least one detection with an \texttt{autoscan} machine learning score of $>0.7$. On each night of observation after the first, we additionally require that each candidate must have an \texttt{autoscan} score of $>0.7$ on multiple nights - two nights for the second epoch, and at least three nights for the third and fourth epochs. We also require that the candidates fall within the 90\% confidence area of the LVK skymap. After initial selection, candidates are visually inspected by human experts for remaining imaging artifacts, bad quality, or improper image subtractions by the processing pipeline before being reported to the astronomical community in a GCN.

Candidates were labeled as nuclear (AGN-like) or supernova-like based on their proximity to their host galaxy (see Table \ref{tab:desgw_candidates}).

\begin{table*}
\hskip-1.5cm\begin{tabular}{|c|c|c|c|c|c|}
    \hline
    Epoch & Initial Candidates & \texttt{autoscan} $>0.7$ & Multiple Detections & Within 90\% Region & Visual Inspection \\
    \hline
    1 & 4103 & 816 & 816 & 598 & 20 \\
    \hline
    2 & 15858 & 1738 & 443 & 281 & 25 \\
    \hline
    3 & 20474 & 2477 & 129 & 80 & 26 \\
    \hline
    4 & 38448 & 3955 & 459 & 289 & 3 \\
    \hline
\end{tabular}
\caption{The DESGW candidate vetting process for each night of observation, showing the number of candidates that pass each listed criterion from left to right. Because only one night of observing had been done after epoch 1, no candidates in epoch 1 failed the detection criterion.}
\label{tab:desgw_candidates}
\end{table*}

\subsection{PFS Data Reduction and Spectral Classification}\label{subsec:PFS_Reduction}

After our PFS observations described in Section \ref{subsec:PFS_observations}, we reduce the data with the PFS Data Reduction Pipeline \citep{Tamura2024}. The reduction procedures include removal of basic instrumental signatures, wavelength calibration, extraction of spectra, sky subtraction, merging spectra from three spectral arms, flux calibration, and coadding multiple exposures. The final products are wavelength-calibrated, sky-subtracted, flux-calibrated, and properly masked one-dimensional science spectra of our 3897 targets. In the reduced spectra, wavelengths masked by, e.g., bad pixels, saturation, or cosmic rays, were not used in the analysis hereafter.

After obtaining the spectra of our 3897 targets, we perform spectral classification, including spectral type and redshift measurements. We used \texttt{Redrock}, a spectral fitter developed and used by the Dark Energy Survey Instrument (DESI) survey \citep{Levi2013,DESI2016,Anand2024}, to obtain the initial spectral type and redshift solutions. The \texttt{Redrock} solutions were selected from the least chi-square templates based on principal component analysis (PCA) and physical galaxy models (archetypes). The templates included three main spectral types of galaxy, star, and QSO. The typical error of our redshift measurements is $<0.001(1+z)$.

After initial spectral classification with Redrock, we perform visual inspection (VI) of all of our 3897 spectra. During the VI, we use a supernova identification tool named \texttt{SNID} \citep{Blondin2007} to measure the supernova types and redshifts for targets we found with potential signatures of supernovae. The data of the PFS NIR arm is not used in the spectral classification from Redrock and SNID, because the NIR data were much noisier than those of blue and red arms (see Figure \ref{fig:SN_QSO_galaxy}). VI is performed independently and cross-checked by two trained experts. We adopt a scoring system to evaluate how well the classification is for each target. The VI scores ranged from A to E, ordered from most to least confident. For targets with VI scores of D and E, we classified them as UNCLEAR type and redshift.

The criteria and number of targets for each VI score are presented in Table \ref{tab:vi}. For our analysis hereafter, we only use targets with VI scores of A and B.

\begin{table*}
\hskip-0.5cm\begin{tabular}{|c|c|c|}
    \hline
    Score & Criteria & Number of Targets \\
    \hline
    A & At least two secure spectral features & 3093 \\
    \hline
    B & At least one secure spectral feature with multiple weak features & 147 \\
    \hline
    C & One strong spectral feature but without other features to confirm what it is  & 37 \\
    \hline
    D & Clear signal but without identified features or contaminated by nearby bright stars & 167 \\
    \hline
    E & No signal & 453 \\
    \hline
\end{tabular}
\caption{Criteria and number of targets of VI scores based on our PFS spectra. For our analysis, we only use targets with VI scores of A and B. }
\label{tab:vi}
\end{table*}

\section{Results and Discussion} \label{sec:results}

\subsection{DECam Transient Candidates}\label{subsec:DECAMCandidates}

Using data obtained from our four epochs of DECam observations, we identified 36 high-confidence transient candidates.

From the first epoch of observations, DESGW reported 20 high-confidence candidates in GCN 39934 \citep{GCN39934}, including:
    \begin{itemize}
        \item Eight nuclear candidates (likely AGNs)
        \item Eight supernovae candidates
        \item Three uncategorized candidates
        \item One candidate identified as known AGN \\ (WISEA J093710.05+082057.2)
    \end{itemize}
    
From the second epoch of observations, DESGW reported 25 high-confidence candidates in GCN 39992 \citep{GCN39992}, including:
    \begin{itemize}
        \item Fifteen nuclear candidates
        \item Ten supernovae candidates
    \end{itemize}
12 candidates were recovered from the first GCN, and none of the reported Swift-XRT candidates were recovered.
    
After the third and fourth epochs of observation, 15 candidates from the second GCN are ruled out by their spec-z obtained from PFS. Another 8 candidates did not exhibit further transient activity and were thus excluded. None of the Swift-XRT candidates were recovered and none of the additional QSOs reported by J-GEM exhibited any evidence of transient activity. DESGW reported 3 high-confidence candidates in GCN 40455 \citep{GCN40455}. This includes:
    \begin{itemize}
        \item Two SN-like candidates outside of J-GEM footprint
        \item One SN-like candidate unique to the final epoch
    \end{itemize}

\begin{figure}[bh]
    \centering
    \includegraphics[width=\linewidth]{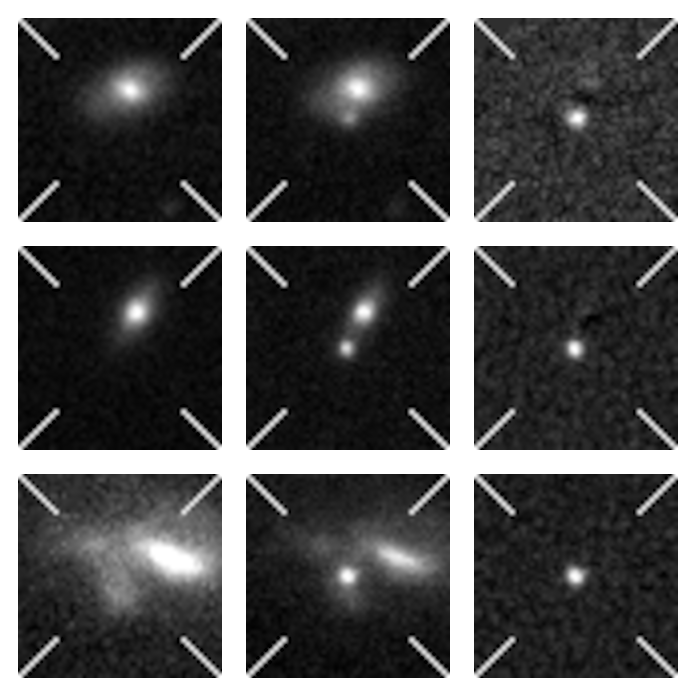}
    \caption{Difference images of the three final DESGW candidates. Each row shows, from left to right, one candidate's original template image, the search image, and the difference, with the transient centered in frame. From top to bottom, the candidates are AT2025kjv, AT2025gem, and AT2025avp. These images have an edge length of around 13.4 arcseconds.}
    \label{fig:DESGW_final_candidates}
\end{figure}

No additional observations were made of the three final candidates reported in the final DESGW GCN. All three of these candidates are clearly separated from their host galaxy by more than 1 arcsecond (see Figure \ref{fig:DESGW_final_candidates}). Because we assume an AGN-driven optical counterpart model for BBH emission, these candidates are likely too separated from the centers of their host galaxies to be related to S250328ae. This means that there are no DESGW candidates that are favored to originate from S250328ae.

\subsection{Example Light Curves and Spectra of SNe, QSOs, and Galaxies}

\begin{figure*}[]
    \centering
    \includegraphics[width=0.85\linewidth]{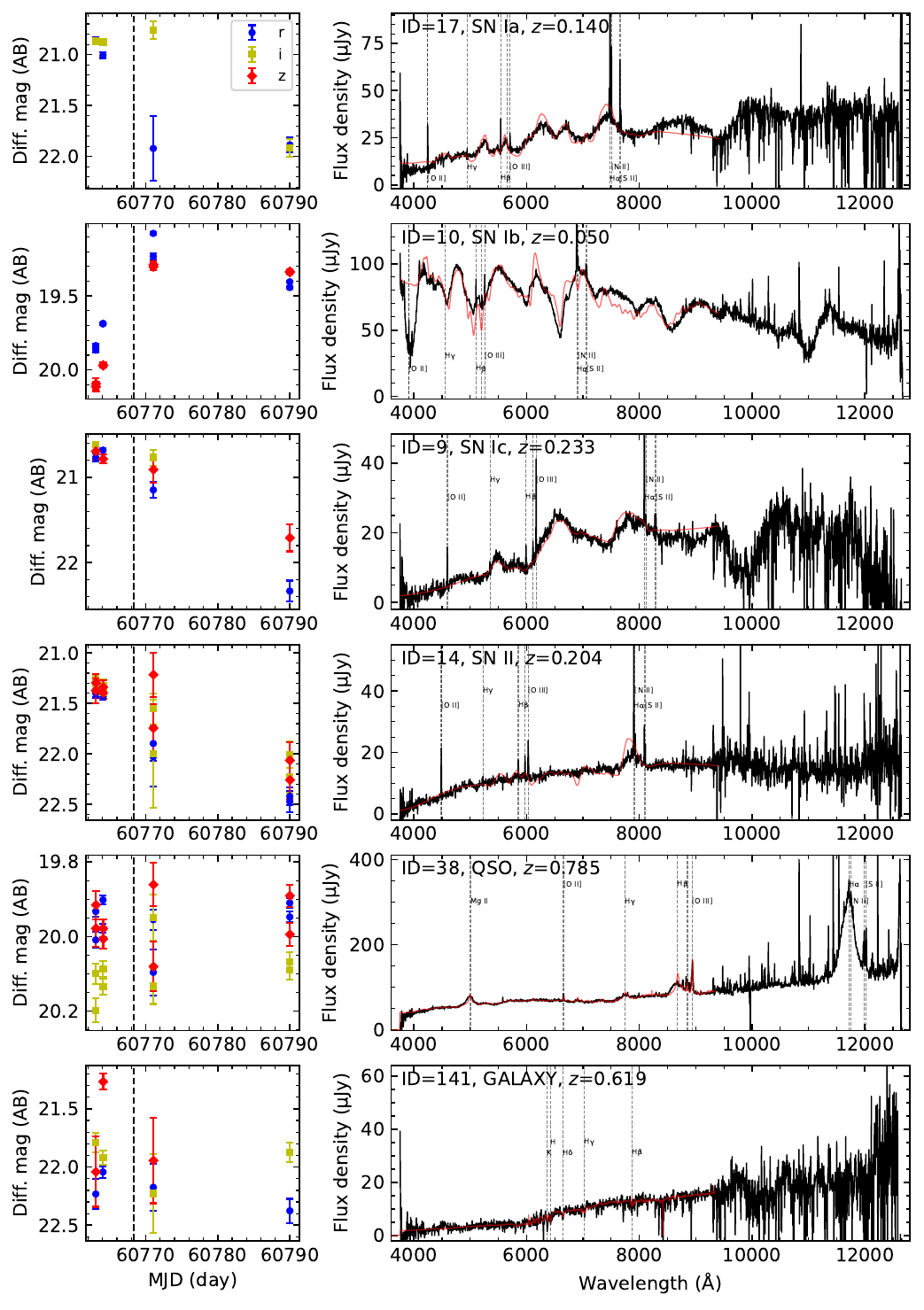}
    \caption{Light curves (left) and spectra (right) of six example SNe, QSO, and galaxy obtained with our DECam and PFS observations. The difference magnitudes of $r$ (blue), $i$ (yellow), and $z$ (red) bands are shown in the left panels as a function of observation dates in MJD. The vertical dashed lines indicate the observation date of PFS. In the right panels, PFS spectra are shown in black and SNID (Redrock) models are shown in red for SNe (QSO and galaxy). Common emission or absorption lines are indicated by dashed vertical lines. For visual clarity, the PFS spectra are binned by 5 {\AA} with the inverse square of noise as the weight.}
    \label{fig:SN_QSO_galaxy}
\end{figure*}

With our DECam and PFS observations, we confirm 12 SNe after visual inspection of spectra. Example light curves and spectra are presented in Figure \ref{fig:SN_QSO_galaxy}. The 12 SNe are all identified as transient candidates by DECam observations. The faintest difference magnitude of 12 SNe is $\sim 23$ mag at the second and third DECam epochs (closest to our PFS observations).

Additionally, we confirm 159 QSOs, 2975 galaxies, and 131 stars with PFS spectra. Among these targets, 139 QSOs, 37 galaxies, and 2 stars are identified as transient candidates with DECam observations.

Example light curves and spectra of these candidates are shown in Figure \ref{fig:SN_QSO_galaxy}. Notably, some QSOs show broad emission lines such as H$\alpha$ and H$\beta$ at near-infrared wavelengths, which are visible in the PFS spectra.

\begin{figure*}[]
    \centering
    \includegraphics[width=0.85\linewidth]{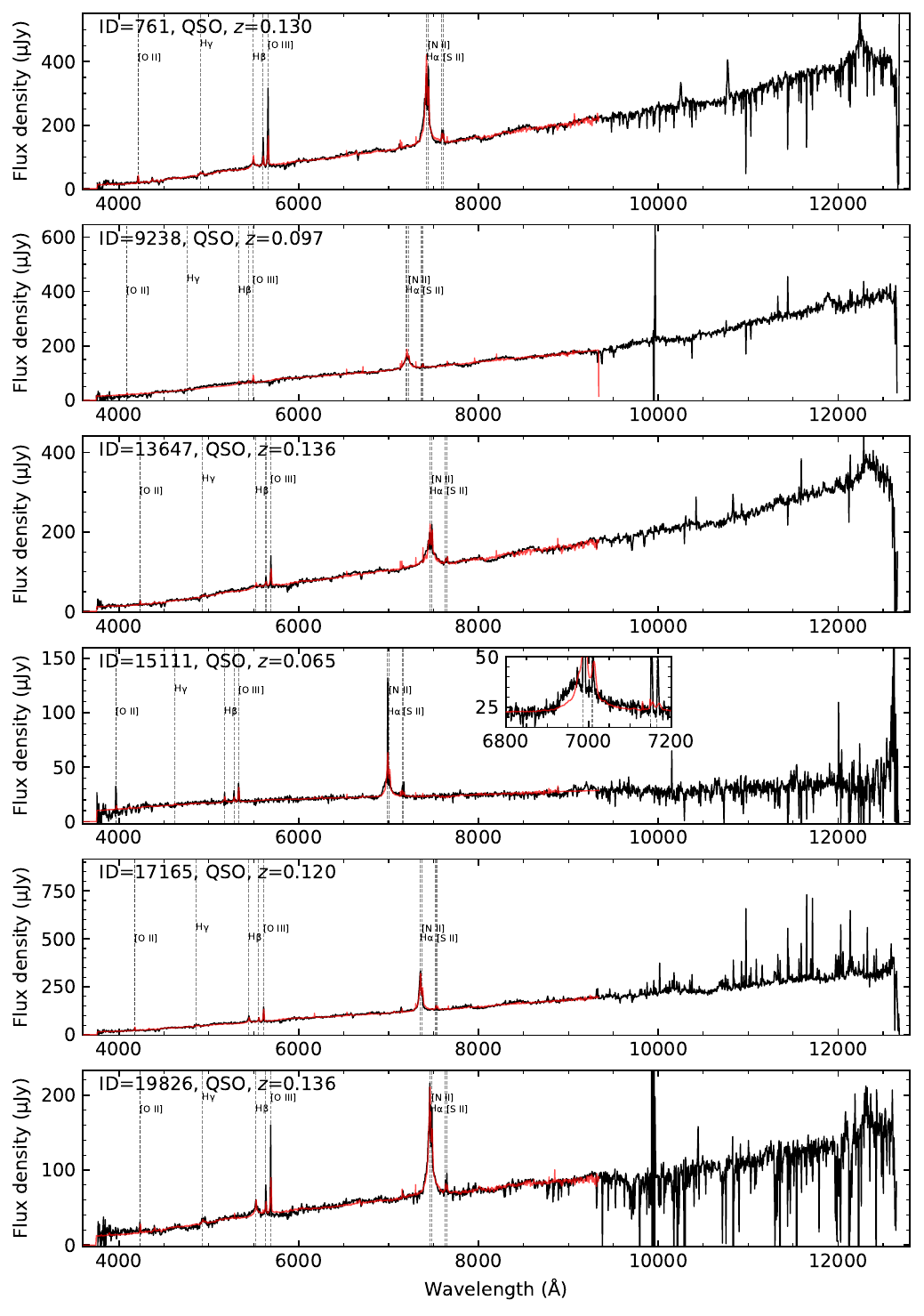}
    \caption{Same as Figure \ref{fig:SN_QSO_galaxy}, but for six QSOs potentially related to S250328ae (within the $\sim 90\%$ probability volume). As these QSOs do not show clear variability in DECam observations, light curves are not shown in this figure. For visual clarity, the PFS spectra (except for the zoom-in panel) are binned by 5 {\AA} with the inverse square of noise as the weight. Although an asymmetric broad H$\alpha$ emission line with a red-shifted tail is identified in one QSO (ID=15111; see the spectrum without binning in the zoom-in panel), we cannot confirm that the feature is caused by asymmetric flare illumination produced by a kicked BBH merger with a single epoch PFS observation.}
    \label{fig:QSO_z0.1}
\end{figure*}

\subsection{Redshift Distribution of Transient Candidates}

\begin{figure}[th]
    \centering
    \includegraphics[width=\linewidth]{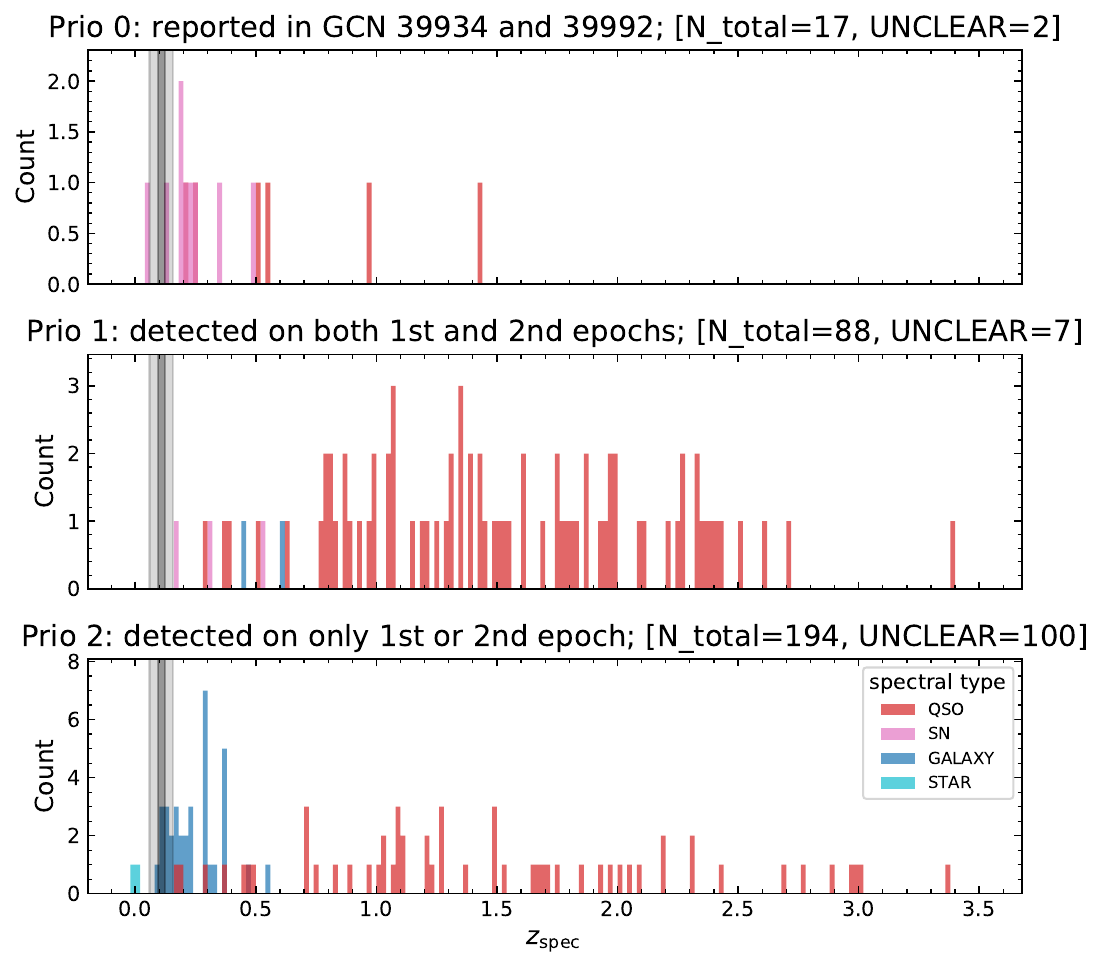}
    \caption{PFS spec-z distribution of DECam transient candidates based on priorities. The total number of transient candidates and UNCLEAR (VI scores of D and E) targets in the priority category is shown in each panel title. The colors of histograms are different by spectral types as shown in the legend. The dark (light) grey shaded region indicates the $1\sigma$ ($3\sigma$) redshift confidence interval of S250328ae assuming the Gaussian distribution.}
    \label{fig:PFSspeczdist}
\end{figure}

Figure \ref{fig:PFSspeczdist} displays the redshift distribution of transients grouped by their assigned priorities.
Since redshift or distance information was not considered in the transient selection process, the spectroscopically confirmed transients are distributed across a wide range of redshifts. These spectroscopic measurements help break the line-of-sight degeneracy inherent in the selection.

A few transients fall within the redshift range of interest. For example, AT 2025cvb is a supernova at 
$z=0.1397$ categorized under Priority 0. Although the explosion dates of these transients are uncertain, their discoveries occurred well before the reported event date.

There are 2/17=12\%, 7/88=8\%, and 100/194=52\% UNCLEAR transients whose spectrum determinations were poor in priorities 0, 1, and 2, respectively. Priorities 0 and 2 include transients that have only one detection from the first two DECam epochs, which potentially include asteroids as well as artifacts that are nominally filtered by the vetting process or visual inspection. The nature of 7 UNCLEAR transients in Priority 1 is not clear. Additionally, a large number of Priority 2 targets are identified as galaxies, because they may be mostly false transients from image subtraction or short time transients. 

\subsection{Potential Candidates Related to S250328ae}

In addition to the candidates highlighted by DESGW, six QSOs are identified by Subaru/PFS within the $\sim 90\%$ probability volume (90\% confidence area with 3$\sigma$ redshift confidence interval) of S250328ae. The spectra of these QSOs are shown in Figure \ref{fig:QSO_z0.1}. Although the six QSOs are not DECam transient candidates and do not show clear variability in DECam observations, we cannot completely rule out the possibility that they may be related to S250328ae, as the EM emission of BBH merger events may be faint. The $3\sigma$ limiting magnitudes of variability for detected transients in the DECam observations are $r\sim23.8$, $i\sim23.5$, and $z\sim23.4$. Five of the six QSOs have been reported in GCN 40221 \citep{Zhang2025GCN}. 

On the other hand, if an off-center disk flare is produced by a kicked BBH merger, the flare will illuminate the clouds in the broad-line region of AGNs asymmetrically. As a result, it is expected that the spectra of EM counterparts in this scenario show asymmetric broad line profiles within a few weeks after events \citep[][see Figure \ref{fig:DESGW_BBHLC}]{McKernan2019,Graham_2020}. Although we identify an asymmetric broad H$\alpha$ line with a red-shifted tail in one of the six QSOs (ID=15111; RA=144.843150 and Dec=11.214702 in degree), we cannot confirm that the asymmetry is caused by flare illumination with a single epoch of PFS observations. Further spectroscopic follow-up observations are encouraged. We do not identify clear asymmetric broad emission lines on the spectra of the other five QSOs. 

In summary, we do not find conclusive evidence suggesting that the six QSOs are related to S250328ae, although we cannot completely rule out the possibility if the EM counterpart of S250328ae is faint.

\subsection{Comparison between PFS and Archival Redshifts}

\begin{figure}[th]
    \centering
    \includegraphics[width=\linewidth]{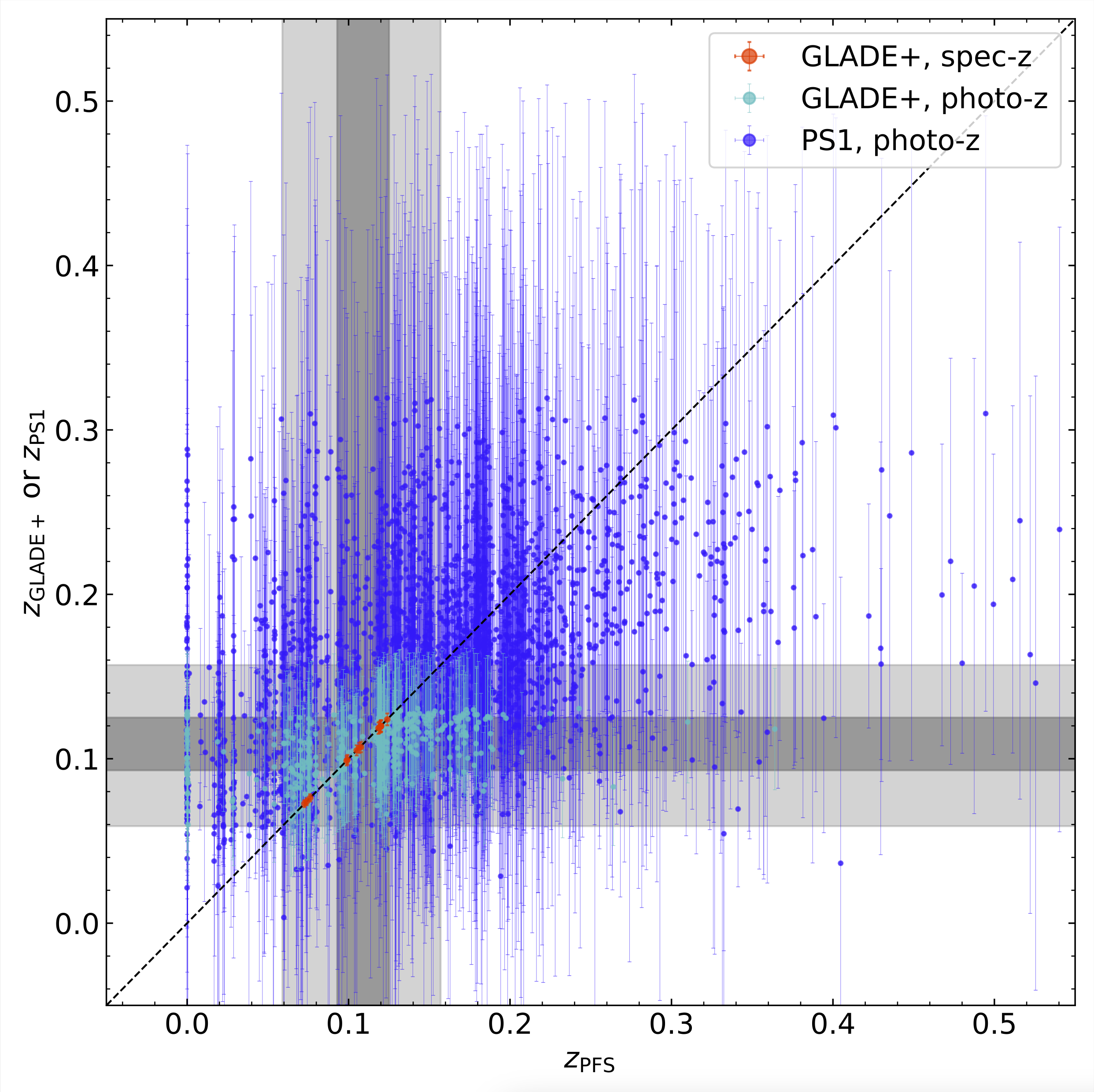}
    \caption{GLADE+ and PS1-STRM redshifts as a function of PFS spec-z. The dots represent PFS targets with GLADE+ spec-z (red), GLADE+ photo-z (cyan), and PS1-STRM photo-z (blue). The dark (light) grey shaded region indicates the $1\sigma$ ($3\sigma$) redshift confidence interval of S250328ae assuming the Gaussian distribution ($511\pm82$ Mpc). The shaded region is just an estimation as the actual redshift interval varies depending on coordinates. The dashed diagonal line shows where the PFS and archival redshifts are the same. The horizontal distribution of GLADE+ and PS1 targets is due to our target selection criteria of redshift. Because the typical error of PFS spec-z is very small ($<0.001(1+z)$), the horizontal error bars are almost invisible. For visual clarity, 27 targets with $z_{\rm PFS}>0.55$ are not shown in this figure.}
    \label{fig:z_compare}
\end{figure}

We compare our PFS spec-z (VI score of A and B) with archival (GLADE+ and PS1-STRM) redshifts, as shown in Figure \ref{fig:z_compare}. The numbers of targets with GLADE+ spec-z, GLADE+ photo-z, and PS1-STRM photo-z are 14, 464 and 2572, respectively. Among these targets, 14, 442, and 2255 (100\%, 95\%, and 88\%) targets have redshift differences of $<0.1(1+z)$, respectively. All of the 14 targets with GLADE+ spec-z have exactly matched PFS spec-z with redshift differences of $<0.001(1+z)$. 

\subsection{Future Perspective}\label{subsec:future}

In this study, we only observed potential targets related to a BBH merger event (S250328ae). The same observation strategy and techniques can also be applied to other GW-related events like kilonovae (e.g. \citealt{GW170817}). Subaru/PFS is able to detect the EM emission of kilonovae at distances of $>200$ Mpc with apparent magnitudes of $\gtrsim 22$ mag (by binning $\sim 4$ pixels; $\sim 4\times 0.8 = 3.2$ {\AA}) within 3-4 days after events (\citealt{Ohgami2021}). With the large field of view and numerous science fibers of PFS, we will be able to search for the EM counterparts efficiently.

As the PFS observation in this study is the first ToO observation since the science operation of PFS have commenced, based on this experience, improvements will be made for future PFS observations. For example, more science targets can be observed with each pointing. For S250328ae, we only observed confident targets such as PS1-STRM galaxies with photo-z errors of $<0.2$, resulting in low fiber usage ($<30\%$; the remaining $>70\%$ fibers were used by observatory fillers). In future PFS observations, we will fill the fibers with low-confidence targets to increase the number of science targets in each pointing.

\section{Conclusion} \label{sec:summary}

In this study, we performed DECam and PFS observations to jointly search for the EM counterpart to the GW BBH merger event S250328ae. Our results are summarized as follows.

\begin{enumerate}
    \item We carried out four epochs of DECam observations covering $>90\%$ of the 90\% confidence area and reported 36 high-confidence transient candidates. Including both high-confidence and lower-confidence candidates, a total of 299 transient candidates were observed by follow-up spectroscopic observations with Subaru/PFS. 

    \item We carried out Subaru/PFS observations to obtain spectra of DECam transient candidates, Swift-XRT candidates, and potential host galaxies of S250328ae. A total of 3897 targets were observed by seven PFS pointings covering $\sim 50\%$ of the 90\% confidence area. After template fitting and careful visual inspection, we identified 12 SNe, 159 QSOs, 2975 galaxies, and 131 stars.
    
    \item With the joint observations of DECam and PFS, we found variability in 12 SNe, 139 QSOs, 37 galaxies, and 2 stars. By positive spectral identification and spectroscopic redshift measurement, all objects that exhibited variability and were observed by both instruments were determined not to be related to the GW event. The three variable candidates reported by DESGW in \cite{GCN40455} have not been ruled out but are likely to be supernovae because of their separation from their host galaxy's nucleus.

    \item Within the $\sim90\%$ localization volume of S250328ae, although no variable QSOs were found, we identified six QSOs without clear variability. If the EM counterpart of S250328ae is faint, these six QSOs may be related to this BBH merger event. Although an asymmetric broad H$\alpha$ emission line with a red-shifted tail is identified in one QSO, we cannot confirm that the feature is caused by asymmetric flare illumination produced by a kicked BBH merger with a single epoch PFS observation.

    \item Based on our observations in this study, we plan to carry out additional PFS observations to cover the entire 90\% confidence area of S250328ae. Using the PFS spec-z across the entire 90\% volume, we expect to accurately estimate the Hubble constant with the dark siren technique (e.g. \citealt{Ballard2023}), and present the results in a future paper.

\end{enumerate}

The Vera C. Rubin Observatory's Legacy Survey of Space and Time (LSST) is set to begin its ten-year survey in late 2025 in Cerro Pachón, Chile. ToO observations are already planned as part of the LSST \citep{2024arXiv241104793A}. The powerful combination of an 8.4-meter aperture, a 3.5-degree diameter field of view, and a high imaging cadence will enable the discovery of a large number of candidate counterparts during Rubin Observatory's ToO observations. While discovery is the primary focus of the Rubin Observatory ToO strategy, the characterization of EM counterparts remains a critical priority for the astrophysical community. 

Multi-object spectroscopy with PFS will be among the most efficient ways to follow up LSST transient candidates. The combination of LSST's discovery capabilities and PFS's spectroscopic follow-up will open a new era of rapid and comprehensive EM counterpart classification. 

Wide-field imagers like DECam and Subaru/HSC will maintain an important role in characterizing the transients by obtaining light curves for confirmed EM counterparts. This includes long-term observations of ToOs that Rubin does not continue observing after discovery, and northern hemisphere ToOs in the case of HSC. DECam and HSC also possess the unique ability to complement Rubin observations by using medium and narrow band filters, which provide additional characterization information about EM counterparts. In addition to these characterization capabilities, DECam and HSC can search for ToOs that do not pass Rubin Observatory's alert criteria, such as binary-black hole GW events with a 90\% confidence area $>20\deg^2$. 

This collaboration and observing strategy lays the groundwork for not only DECam and PFS, but also other mid-size wide-field imagers and multi-object spectrograph observing campaigns in the 2020s and beyond. We show that by combining observations with wide-field imaging instruments and multi-object spectrographs, we are able to effectively search  for EM counterparts of GW events and classify hundreds of transients, independent of the discovery of an associated EM counterpart.

\newpage
\begin{acknowledgments}

This document was prepared by the Dark Energy Survey Gravitational Wave (DESGW) Collaboration using the resources of the Fermi National Accelerator Laboratory (Fermilab), a U.S. Department of Energy, Office of Science, Office of High Energy Physics HEP User Facility. Fermilab is managed by FermiForward Discovery Group, LLC, acting under Contract No. 89243024CSC000002.

The DECam Search \& Discovery Program for Optical Signatures of Gravitational Wave Events (DESGW) is carried out by the Dark Energy Survey (DES) collaboration in partnership with wide-ranging groups in the community. DESGW uses data obtained with the Dark Energy Camera (DECam), which was constructed by the DES collaboration with support from the Department of Energy and member institutions, and utilizes data as distributed by the Science Data Archive at NOIRLAB. NOIRLAB is operated by the Association of Universities for Research in Astronomy (AURA) under a cooperative agreement with the National Science Foundation. We thank the Cerro Tololo observatory staff for their support in acquiring these observations.

This work is supported by JSPS KAKENHI Grant Numbers 24H00027, 24K17097, 23H04894, and 25K07361.

We thank all those who contributed to the PFS project, including collaborators, technical and administrative staff, and institutional partners.
The PFS project has been supported by the WPI Initiative, MEXT, and by funding agencies in Japan (JSPS), the US (NSF), Taiwan (Academia Sinica), Brazil (FAPESP, CNPq), and France (CNRS, Aix Marseille University).

Part of the data analyses presented in this paper were carried out at the Prime Focus Spectrograph Science Platform, which is operated by the Subaru Telescope and the Astronomy Data Center at the National Astronomical Observatory of Japan.

This work is based in part on software developed by the Dark Energy Survey Instrument (DESI) survey. We appreciate the DESI Collaboration for their contributions and publicly sharing of the DESI software.

This work is based in part on data collected at Subaru Telescope, which is operated by the National Astronomical Observatory of Japan. We are honored and grateful for the opportunity of observing the Universe from Maunakea, which has the cultural, historical, and natural significance in Hawaii.

\end{acknowledgments}

\begin{contribution}

HZ, MK, NT, and YU: Mainly contributed to sections \ref{sec:intro}, \ref{subsec:PFS_observations}, \ref{subsec:PFS_Reduction}, \ref{sec:results}, and \ref{sec:summary}.

SM, IM, LJ, and SK: Mainly contributed to sections 
\ref{sec:intro}, \ref{sec:Data}, \ref{subsec:DESGW candidate vetting}, \ref{subsec:DECAMCandidates}, and \ref{sec:summary}.

The other authors contributed to DECam and PFS observations, data reduction, data analysis, and manuscript preparation.


\end{contribution}

%
\facilities{Subaru Telescope:8.2m, Prime Focus Spectrograph}
\facilities{CTIO Blanco:4m, Dark Energy Camera}

\software{Main-Injector \citep{MIPaper}, SExtractor \citep{SourceExtractor}, scamp \citep{scamp}, diffimg \citep{Kessler_2015}, autoscan \citep{AutoscanPaper}, Redrock \citep{Anand2024}, SNID \citep{Blondin2007}}






\bibliography{sample7}{}
\bibliographystyle{aasjournalv7}



\end{document}